\documentclass[conference]{IEEEtran}
\IEEEoverridecommandlockouts

\usepackage{cite}

\usepackage{multirow}
\usepackage{makecell}

\usepackage[ruled,vlined,linesnumbered]{algorithm2e}

\usepackage[colorlinks=true, linkcolor=blue, citecolor=blue, urlcolor=magenta]{hyperref}
\usepackage{array}
\usepackage[caption=false,font=normalsize,labelfont=sf,textfont=sf]{subfig}
\usepackage{textcomp}
\usepackage{stfloats}
\usepackage{url}
\usepackage{verbatim}
\usepackage{graphicx}
\usepackage{epstopdf}
\usepackage{dsfont}
\usepackage{amssymb}
\usepackage{amsthm}
\usepackage{amsmath}

\ifCLASSINFOpdf
\else
\fi

\begin{document}

\title{EH-FedSAG: Variance-Reduced Federated Learning with Energy-Aware Participation in Energy-Harvesting IoT}

\author{\IEEEauthorblockN{Shahab Jahanbazi\IEEEauthorrefmark{1},
Mateen Ashraf\IEEEauthorrefmark{1}, 
Richard Demo Souza\IEEEauthorrefmark{2} and
Onel~L.~A.~L\'opez\IEEEauthorrefmark{1}}
\IEEEauthorblockA{\small{\IEEEauthorrefmark{1}Centre for Wireless Communication Systems (CWC), University of Oulu, Oulu, Finland}}
\IEEEauthorblockA{\small{\IEEEauthorrefmark{2}Electrical and Electronics Engineering Department, Federal University of Santa Catarina, Florian\'opolis, Brazil}}
\IEEEauthorblockA{\small{Emails: \{Shahab.Jahanbazi, Mateen.Ashraf, Onel.AlcarazLopez\}@oulu.fi, richard.demo@ufsc.br}}
\thanks{This work is supported by the Research Council of Finland (Grants 362782 (ECO-LITE), and 369116 (6G Flagship)), the European Commission through the Horizon Europe/JU SNS project AMBIENT-6G (Grant 101192113), CNPq INCT STREAM (409179/2024-8) and RNP/MCTI Brasil 6G project (01245.020548/2021-07).}}

\maketitle

\begin{abstract}
Federated learning (FL) in energy-harvesting (EH) networks is challenged by intermittent and stochastic energy arrivals that lead to unstable device participation across training rounds, and by high communication costs under limited energy budgets, reducing overall training efficiency. This paper studies FL under a slot-based EH model and proposes EH-FedSAG, a server-memory-based variance-reduced method. We compare EH-FedSAG with vanilla EH-FedAvg under the same multi-channel orthogonal multiple-access uplink model and within a unified simulation framework that captures battery charging, local computation cost, and transmission cost under different energy-arrival probabilities. Performance is assessed in terms of test accuracy over training rounds for both homogeneous and heterogeneous data distributions. The results show that EH-FedSAG consistently achieves higher test accuracy than EH-FedAvg in the considered settings, while exhibiting substantially lower training variance. The advantage of EH-FedSAG is more pronounced under scarce energy availability and non-independent/identically-distributed data.
\end{abstract}


\section{Introduction}
\IEEEPARstart{F}{ederated} learning (FL) enables edge devices to collaboratively train a shared model without transmitting raw data to a central server, making it well suited to privacy-sensitive and bandwidth-limited applications~\cite{mcmahan}. Among FL methods, FedAvg is the standard baseline. In each FedAvg communication round, participating devices perform local stochastic gradient updates, and the server aggregates the resulting model parameters by averaging. Due to unreliable connectivity and local/network resource constraints full device participation in every communication round is often infeasible~\cite{FedPAQ, FedAvg_non}. As a result, only a subset of devices participates in each round, and this subset is frequently formed stochastically in baseline approaches~\cite{mcmahan, random_participation}, while many works investigate more advanced client selection strategies~\cite{Client_selection, client_selection2}. The performance of FedAvg depends strongly on both the set of participating devices and the communication conditions under which they operate. In particular, partial participation introduces variability in the aggregated updates, which leads to a non-vanishing error floor in the convergence behavior of FedAvg~\cite{FedVARP}. To mitigate this effect, server-side variance-reduction techniques have been proposed, such as FedVARP~\cite{FedVARP}. These approaches are conceptually related to incremental gradient methods, such as SAGA~\cite{SAGA}, which leverage memory to construct lower-variance update directions.

On the other hand, the increasing deployment of networks composed of energy-constrained devices has motivated the study of energy-harvesting (EH) FL systems~\cite{Rafael1, energy_harvesting_FL, battery-aware}. In such systems, unlike conventional edge devices with stable power supplies, devices rely on intermittent and stochastic energy sources, leading to time-varying energy availability. Since these EH devices exhibit dynamically evolving energy buffer states, their capability to participate in the training process is affected. Indeed, device participation in EH-FL is constrained not only by communication and system factors, but also by the availability of sufficient energy for local computation and uplink transmission. This additional energy constraint introduces an additional layer of randomness, further exacerbating the partial participation inherent in conventional FL systems.
Prior works on EH-FL have therefore focused on energy-aware device selection~\cite{Rafael1, energy_harvesting_FL} and energy-efficient protocol design~\cite{battery-aware, FedSeq}. These studies demonstrate that time-varying energy availability further amplifies partial participation, significantly impacting convergence behavior~\cite{energy_harvesting_FL, battery-aware, Hamdi_2022}.

Along these research directions, the combined impact of EH constraints, energy-aware participation, and server-side memory-based variance reduction has not yet been studied in a unified framework. In particular, it is not yet clear how server-memory-based variance reduction should be incorporated into an EH-FL protocol in which slot-level battery evolution, training/transmission feasibility, and limited uplink resources jointly determine participation in each round. Motivated by this gap, we consider an EH-FL system under a slot-based EH model, where local training and uplink transmission consume explicit energy units and occupy dedicated time slots. We propose an EH-FL scheme based on the SAGA principle, termed EH-FedSAG, in which devices with sufficient energy upload their local model updates to the server through a set of orthogonal channels. Consequently, in each communication round, only a limited number of devices can participate. Under this common setting plus identical energy constraints, we compare the conventional EH-FedAvg approach (i.e., the vanilla FedAvg procedure adapted to EH systems) with the proposed EH-FedSAG approach.

The main contributions of this paper are threefold. 
First, we propose an energy-aware EH-FL framework that explicitly captures slot-level battery dynamics, employs an age-based device selection mechanism, and enforces a per-round participation constraint.
Second, we develop EH-FedSAG, a memory-based variant of EH-FedAvg that incorporates server-side memory to mitigate variance under time-varying device availability and partial participation.
Third, we characterize, through experiments, the relationship between test accuracy and training rounds under different EH conditions across two different networks. The results show that EH-FedSAG is more impactful when energy arrivals are scarce and when the data are statistically heterogeneous.

\section{System Model}
\label{sec:system_model}
\begin{figure}
\centering
\includegraphics[width=0.45\textwidth]{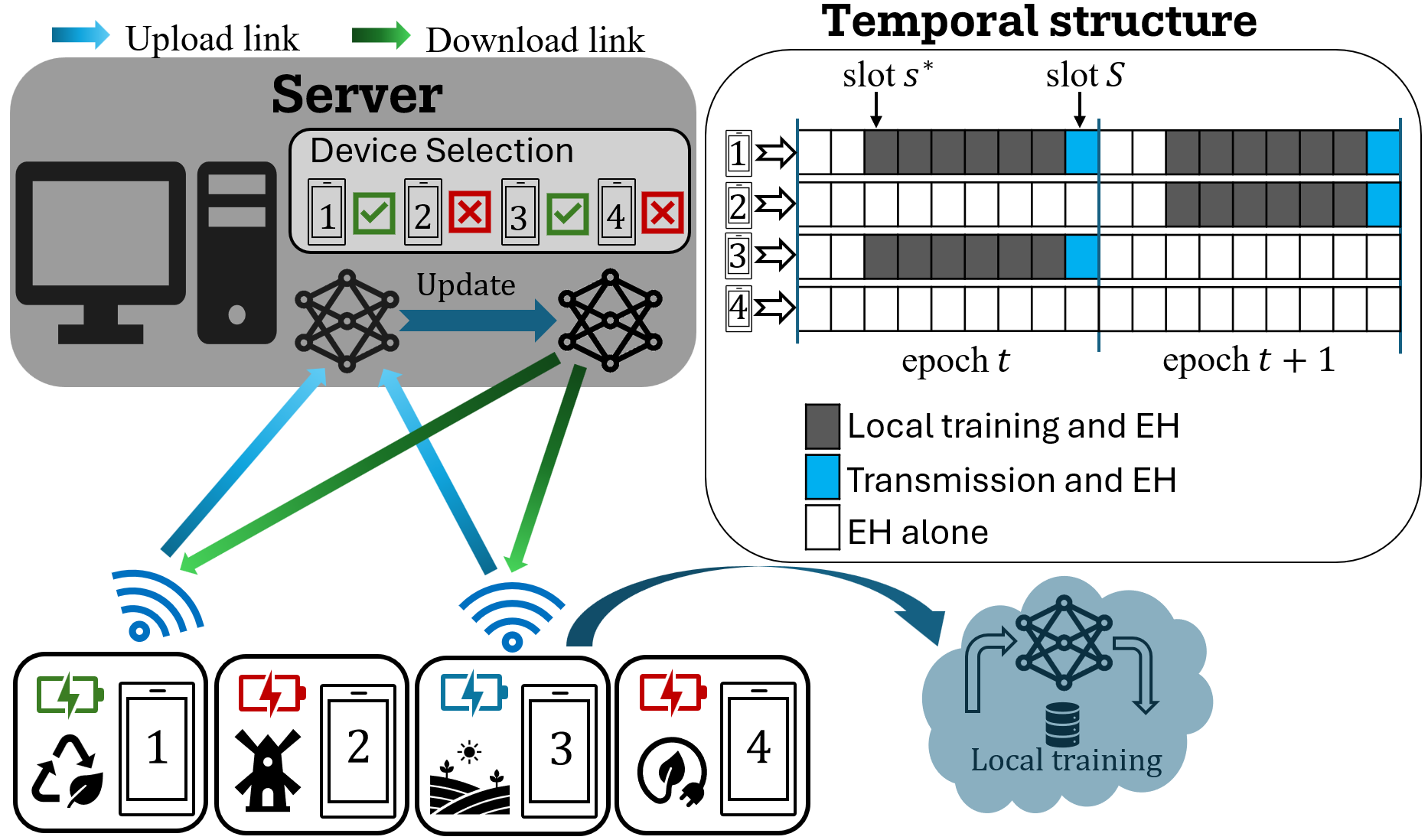}
\caption{System model of the considered scenario, consisting of four EH devices selected to participate in the distributed optimization algorithm based on their available energy levels. The temporal structure of the algorithm follows a synchronous slotted protocol in which devices $1$ and $3$ participate at round $t$, while devices $1$ and $2$ participate at round $t+1$.}
\label{fig:system_model}
\end{figure}

We consider an EH-FL system in which a central server coordinates $K$ distributed devices via a shared wireless uplink, as illustrated in Fig.~\ref{fig:system_model}. The objective of FL is to collaboratively learn a global model parameter vector $\mathbf{w} \in \mathbb{R}^d$ by keeping all training data locally at each device.
Formally, the learning task is cast as the minimization of the global empirical risk
\begin{equation}
    F(\mathbf{w}) = \frac{1}{K} \sum_{k=1}^{K} f_k(\mathbf{w}),
\end{equation}
where $f_k(\mathbf{w})$ denotes the local objective function associated with device $k$, defined as
$f_k(\mathbf{w}) \triangleq \mathbb{E}_{\xi_k \sim \mathcal{D}_k} \big[ \ell(\mathbf{w}; \xi_k) \big]$.
Here, $\ell(\cdot;\cdot)$ represents the loss function evaluated on a data sample, and $\mathcal{D}_k$ denotes the underlying data distribution at device $k$.

The learning process is coordinated by the server through iterative communication rounds. At each round, the server distributes the current global model to participating devices. Each device then performs local training using its dataset and subsequently transmits the resulting model update to the server. The server aggregates the received updates to obtain the new global model for the next round. To model the temporal structure of this process, we adopt a synchronous slotted FL protocol in which communication proceeds in rounds indexed by $t\in \mathbb{Z}^+$.\footnote{The purpose of this model is to create EH-driven intermittency as finite battery storage, stochastic energy arrivals, explicit energy consumption for local training and transmission, and a per-round limit on the number of devices that can upload.} Each round is divided into $S$ equal-duration time slots indexed by $s\in\{1,\ldots,S\}\triangleq\mathcal{S}$.

\subsection{Energy Dynamics}
Each device is equipped with an EH module and a finite-capacity battery. We assume the harvest-store-use paradigm~\cite{harvest_store_use} in which harvested energy from the environment is first stored and becomes available in subsequent slots. Specifically, let $e_k(t,s)\in\{0,1\}$ denote the energy harvested by device $k$ during slot $s$ of round $t$, and assume an i.i.d.\ Bernoulli EH process across devices, rounds, and slots with $\Pr\!\big(e_k(t,s)=1\big)=P_h$.\footnote{The Bernoulli EH assumption provides a tractable abstraction for intermittent and memoryless energy arrivals in time-slotted systems and is commonly used in prior work (e.g.,~\cite{battery-aware, energy_harvesting_FL}).}
In addition, energy consumption is modeled in discrete units in which a device consumes one unit of energy per active slot and zero units otherwise. Formally, let $c_k(t,s)\in\{0,1,\ldots,C_{\max}\}$ denote the battery level at the beginning of slot $s$ in round $t$, and let $u_k(t,s)\in\{0,1\}$ denote the energy spent by device $k$ in slot $(t,s)$. Therefore, the battery evolution for $k$-th device is given by
\begin{equation}
c_k(t,s{+}1)=\min\Big\{C_{\max},\, c_k(t,s)-u_k(t,s)+e_k(t,s)\Big\},
\label{eq:battery_update}
\end{equation}
for all $s\in\mathcal{S}$, and $c_k(t,S+1) \rightarrow c_k(t{+}1,1)$.

\subsection{Training and Transmission}
In each round, a participating device performs local training over $\Omega$ consecutive active slots and then transmits its local model update to the server, which consumes one additional unit of energy. The last slot of each round is reserved for model transmission. Therefore, local training must be completed by the end of slot $S-1$, and the latest feasible starting slot for training is $s^\star \triangleq S-\Omega$.

To address intermittent power availability and ensure that all participating devices possess sufficient energy to initiate training, the server performs energy-aware device selection. Specifically, devices are selected only if their battery level at $s^\star$ is already sufficient to support both local training and the subsequent transmission, thereby avoiding battery-induced post-training upload failures. Formally, in round $t$, the server first identifies the set of training-feasible devices
\begin{equation}
\mathcal{S}_t \triangleq \Big\{k\in\{1,\dots,K\} \;:\; c_k(t,s^\star)\ge \Omega+1 \Big\},
\label{eq:St_def}
\end{equation}
i.e., devices whose battery level at $s^\star$ is sufficient to complete local training and subsequently transmit its updated information. The device–server uplink consists of $M$ orthogonal channels with $M < K$, allowing at most $M$ devices to be scheduled per round. Accordingly, the server selects a subset $\mathcal{P}_t \subseteq \mathcal{S}_t$ with $|\mathcal{P}_t|=\min\{M,|\mathcal{S}_t|\}$, to participate in local training and transmission.

All in all, under EH-driven partial participation, vanilla EH-FedAvg can suffer from unstable progress because the set of participating devices in a round is governed by channel conditions and stochastic energy availability. To mitigate this issue, we propose EH-FedSAG, an energy-aware server-memory aggregation scheme. The novelty lies in integrating server-side memory into the EH-FL protocol presented in the next section, where fresh updates, stale memories, battery status, and transmission feasibility are jointly handled through the energy-aware device selection participation process.

\section{Proposed Algorithm: EH-FedSAG}
\label{sec:proposed}
As the first step, whenever $|\mathcal{S}_t|>M$, a scheduling rule is required to determine which feasible devices are selected. In this paper, we adopt an age-based policy that promotes long-term fairness by prioritizing devices that have not been selected recently. Specifically, let $\tau_k(t)$ denote the most recent round index before $t$ in which device $k$ was selected, and define its selection age as $a_k(t)\triangleq t-\tau_k(t),$ with $a_k(t)=\infty$ if device $k$ has not been selected previously. The scheduler ranks devices in decreasing order of $a_k(t)$ and selects the top $M$ devices. To implement this policy, the server maintains participation-age counters for all devices and updates them across communication rounds. 

At slot $s^\star$ within round $t$, the server transmits the current global model $\mathbf{w}_t$ to all selected devices. Each participating device $k \in \mathcal{P}_t$ then performs $B$ steps of local SGD, initialized as $\mathbf{w}^{(k)}_{t,0} = \mathbf{w}_t$ and then
\begin{equation}
\mathbf{w}^{(k)}_{t,b+1} = \mathbf{w}^{(k)}_{t,b} - \eta \, \mathbf{g}^{(k)}_{t,b}, \quad b = 0, \ldots, B-1,
\label{eq:local_update}
\end{equation}
where $\eta > 0$ denotes the local learning rate, and $\mathbf{g}^{(k)}_{t,b}$ is a stochastic gradient of the local objective function $f_k(\cdot)$ evaluated at $\mathbf{w}^{(k)}_{t,b}$ using a mini-batch.
Upon finishing local training, device $k$ forms and transmits the update
\begin{equation}
\boldsymbol{\delta}^{(k)}_t \triangleq \mathbf{w}_t - \mathbf{w}^{(k)}_{t,B}.
\label{eq:delta_def}
\end{equation}
The set of successfully received model differences at the server is $\{\boldsymbol{\delta}^{(k)}_t\}_{k\in\mathcal{P}_t}$.

\begin{algorithm}[t]
\caption{EH-FedSAG with Energy-Aware Selection}
\label{alg:eh_fedsag}
\KwIn{Number of devices $K$, round length $S$, training slots $\Omega$, 
local steps $B$, battery capacity $C_{\max}$, EH probability $P_h$, 
learning rates $\eta$ and $\eta_s$}
\KwOut{Global model $\{\mathbf{w}_t\}$}
Initialize global model $\mathbf{w}_1$ and $\mathbf{m}^{(k)}_1=\mathbf{0}, \forall k$\;
\For{$t=1,2,\ldots$}{
At slot $s^\star$, the server selects the set of $\mathcal{P}_t$\;
Server broadcasts $\mathbf{w}_t$ to all devices in $\mathcal{P}_t$\;
\For{$k\in\mathcal{P}_t$ \textbf{in parallel}}{
Perform $B$ local SGD steps starting from $\mathbf{w}_t$ based on~\eqref{eq:local_update} and form $\boldsymbol{\delta}^{(k)}_t$ based on~\eqref{eq:delta_def} \;
Transmit $\boldsymbol{\delta}^{(k)}_t$ to the server \;}

\eIf{$|\mathcal{P}_t|>0$}{
Compute $\bar{\mathbf{m}}_t$ and then $\mathbf{v}_t$ based on \eqref{eq:mean_memory}, \eqref{eq:v_t}\;
Update global model $\mathbf{w}_{t+1}$ based on~\eqref{eq:server_update}\;
\For{$k=1,\ldots,K$}{

\eIf{$k\in\mathcal{P}_t$}{
$\mathbf{m}^{(k)}_{t+1}=\boldsymbol{\delta}^{(k)}_t$\;
}{
$\mathbf{m}^{(k)}_{t+1}=\mathbf{m}^{(k)}_t$\;
}
}
}{
$\mathbf{w}_{t+1}=\mathbf{w}_t$ and $\mathbf{m}^{(k)}_{t+1}=\mathbf{m}^{(k)}_t, \forall k$\;
}}
\end{algorithm}

The server maintains a per-EH device memory vector $\mathbf{m}^{(k)}_t\in\mathbb{R}^d$ storing the most recently received $\boldsymbol{\delta}^{(k)}$.
After receiving updates in round $t$, the server computes the mean memory
\begin{equation}
\bar{\mathbf{m}}_t \triangleq \frac{1}{K}\sum_{k=1}^K \mathbf{m}^{(k)}_t.
\label{eq:mean_memory}
\end{equation}
Then, it forms a variance-reduced direction
\begin{equation}
\mathbf{v}_t \triangleq \bar{\mathbf{m}}_t
+ \frac{1}{|\mathcal{P}_t|}\sum_{k\in\mathcal{P}_t}\Big(\boldsymbol{\delta}^{(k)}_t - \mathbf{m}^{(k)}_t\Big),
\quad \text{if } |\mathcal{P}_t|>0,
\label{eq:v_t}
\end{equation}
and performs the global update
\begin{equation}
\mathbf{w}_{t+1} = \mathbf{w}_t - \eta_s \mathbf{v}_t,
\label{eq:server_update}
\end{equation}
where $\eta_s>0$ is the server learning rate.
Finally, server memories are updated only for participating devices:
\begin{equation}
\mathbf{m}^{(k)}_{t+1}=
\begin{cases}
\boldsymbol{\delta}^{(k)}_t, & k\in \mathcal{P}_t,\\
\mathbf{m}^{(k)}_t, & k\notin \mathcal{P}_t.
\end{cases}
\label{eq:memory_update}
\end{equation}
If no device successfully uploads in round $t$, the server keeps $\mathbf{w}_{t+1}=\mathbf{w}_t$ and $\mathbf{m}^{(k)}_{t+1}=\mathbf{m}^{(k)}_t$. The corresponding procedure is summarized in Algorithm~\ref{alg:eh_fedsag}.
In terms of memory complexity, EH-FedSAG requires maintaining both update-related memory vectors and counter-based memory scalars, leading to a total server memory requirement of $O(K(d+1))$.



\section{Experimental results}
\begin{figure*}
    \centering
    \begin{minipage}{0.92\textwidth}
        \centering
        \includegraphics[width=\linewidth]{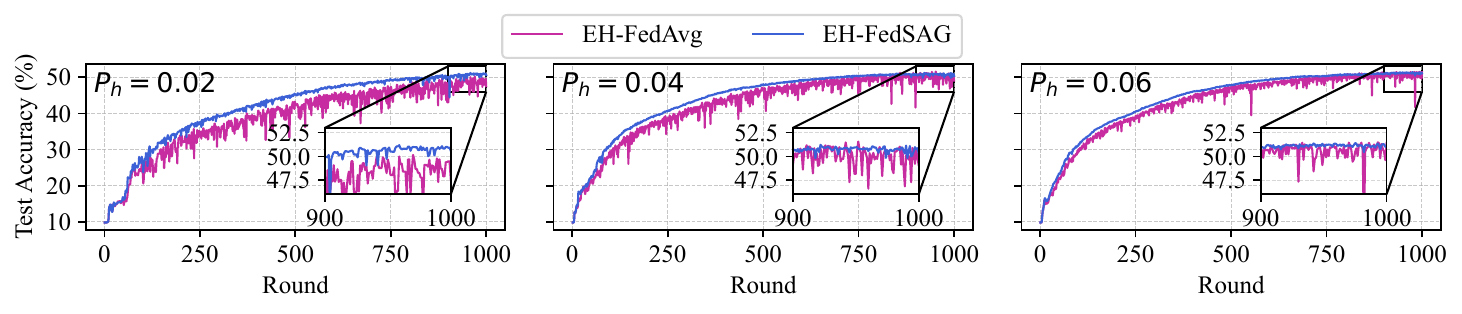}
    \end{minipage}
    \begin{minipage}{0.92\textwidth}
        \centering
        \includegraphics[width=\linewidth]{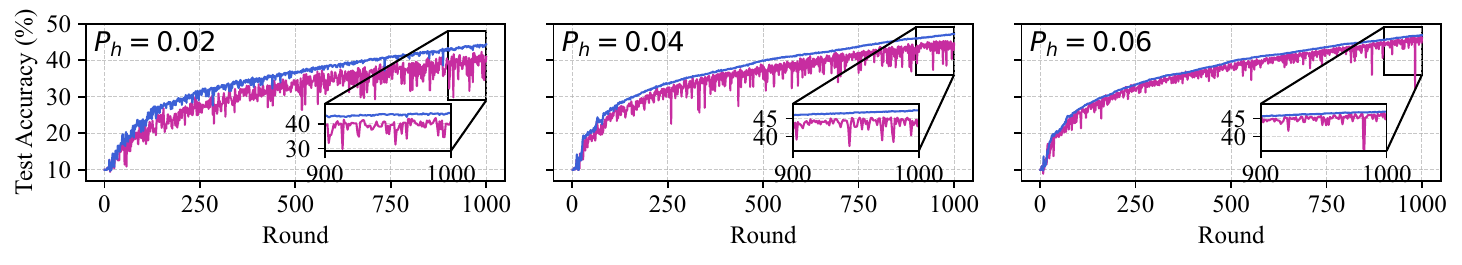}
    \end{minipage}
    \caption{Test accuracy versus communication rounds under different $P_h$ for (top) the SmallCNN network and (bottom) the SmallResNet architecture.}
\label{fig:accuracy_vs_round}
\end{figure*}

To evaluate the proposed EH-FedSAG algorithm, we consider an image classification task on CIFAR-10 and compare it with EH-FedAvg. The EH-FedAvg benchmark follows the same EH model and feasibility constraints. In each round, every participating device performs local SGD steps and transmits its local update $\boldsymbol{\delta}^{(k)}_t$, and the server updates the global model using the simple average of the received updates. The number of devices is $K=100$, the number of orthogonal uplink channels is $M=25$, and training is conducted for $1000$ communication rounds. We consider two networks: (i) a SmallCNN with two convolutional layers, one max-pooling layer, and three fully connected layers; and (ii) a SmallResNet architecture. The local learning rate is set to $0.005$, while the server learning rate for EH-FedSAG is fixed at~$1.0$.
In each round comprising $S=30$ slots, every selected device consumes $\Omega=20$ units of energy to perform $B=5$ local SGD iterations. Finally, the average harvested energy is $30P_h$ units per round and $P_h\in \{0.02,0.04,0.06\}$.
For all considered values of $P_h$, the harvested energy remains significantly smaller than the energy required for one participation round, implying that devices must typically accumulate energy over multiple rounds before becoming active. This results in intermittent and stochastic participation, which is the key operating regime of interest in EH-FL. Moreover, such assumptions are practically relevant for low-power IoT deployments powered by weak ambient energy sources, where the harvested energy over a short interval is often insufficient to support both actions in every round.

Figure~\ref{fig:accuracy_vs_round} shows the test accuracy versus the communication round for EH-FedAvg and the proposed EH-FedSAG under different EH conditions. When $P_h=0.02$, EH-FedAvg exhibits noticeable oscillations due to the severe energy limitation and the resulting stochastic device availability. Nevertheless, EH-FedSAG tends to achieve slightly higher test accuracy (approximately $3\%$ higher) during most of the training process, while also showing improved robustness to participation-induced fluctuations. This behavior can be attributed to the variance-reduction mechanism of EH-FedSAG, which leverages the recently stored information from devices to partially compensate for missing updates from currently unavailable devices.
As $P_h$ increases, the performance of both methods becomes more stable, since more devices are able to accumulate sufficient energy and participate in training more regularly. In this regime, the difference between EH-FedAvg and EH-FedSAG becomes smaller, which is expected: the main advantage of EH-FedSAG lies in mitigating the variance caused by time-varying and stochastic participation, and this source of variance becomes less dominant when energy arrivals are more frequent.

Following~\cite{variance} that treats model performance as a stochastic quantity and analyzes variance in learning behavior, we define a late-stage accuracy variance $V_{\mathrm{late}}$ to capture fluctuations in the converged regime. Let $\text{Acc}_t$ denote the test accuracy at round $t$, and let $\mathcal{T}_{\mathrm{late}}$ denote the final $20\%$ of training rounds. We define $V_{\mathrm{late}}$ as
\begin{equation}
V_{\mathrm{late}} = \frac{1}{|\mathcal{T}_{\mathrm{late}}|} \sum_{t \in \mathcal{T}_{\mathrm{late}}} \left(\text{Acc}_t - \bar{a}_{\mathrm{late}}\right)^2,\\
\end{equation}
where $\bar{a}_{\mathrm{late}} = (1/|\mathcal{T}_{\mathrm{late}}|) \sum_{t \in \mathcal{T}_{\mathrm{late}}} \text{Acc}_t$. 
Lower $V_{\mathrm{late}}$ indicates more predictable model behavior across rounds.
In addition, to quantify convergence speed, we define the \textit{time-to-target} metric
\begin{equation}
T_{\tau}^{(\Xi)} \triangleq \min \left\{ t \;:\; \text{Acc}_{t'} \ge \tau,\ \forall t' \in \{t, t+1, \dots, t+\Xi-1\} \right\},\\
\end{equation}
i.e., the first round at which the test accuracy reaches a target level $\tau$ and stays above it for $\Xi$ consecutive rounds.
\begin{table}[t]
\centering
\caption{Performance comparison of EH-FedAvg and EH-FedSAG under different EH probabilities $P_h$ over the SmallCNN network.}
\label{tab:ph_comparison_metrics}
\begin{tabular}{c l c c c c}
\hline
$P_h$ & Method & $T_{50\%}^{(10)}$ & \makecell[l]{Final \\ Acc. ($\%$)} & \makecell[l]{Best \\ Acc. ($\%$)} & \makecell[l]{Late-stage \\ Acc. variance} \\
\hline
\multirow{2}{*}{$0.02$} & EH-FedAvg & N/A & 47.51 & 50.16 & 2.0915 \\
 & EH-FedSAG & \textbf{915} & \textbf{50.18} & \textbf{51.18} & \textbf{0.5350} \\
\hline
\multirow{2}{*}{$0.04$} & EH-FedAvg & N/A & 49.07 & \textbf{51.56} & 1.5751 \\
 & EH-FedSAG & \textbf{708} & \textbf{50.35} & 51.22 & \textbf{0.0468} \\
\hline
\multirow{2}{*}{$0.06$} & EH-FedAvg & 903 & 51.04 & \textbf{51.48} & 0.9401 \\
 & EH-FedSAG & \textbf{668} & \textbf{51.41} & 51.40 & \textbf{0.0421} \\
\hline
\end{tabular}
\end{table}
Table~\ref{tab:ph_comparison_metrics} complements the learning curves by quantifying final accuracy, best accuracy, convergence speed, and late-stage stability. The table shows that using the time-to-target metric $T_{50\%}^{(10)}$, EH-FedSAG attains a stable test accuracy of $50\%$ with approximately $26\%$ fewer communication rounds than EH-FedAvg for $P_h=0.06$. Across all considered values of $P_h$, while the final accuracy remains comparable, EH-FedSAG consistently achieves significantly lower late-stage variance (e.g., $0.9401$ vs. $0.0421$ for $P_h=0.06$). This combination of faster time-to-target and lower variance suggests that the model reaches and maintains stable performance earlier, which is particularly beneficial in EH IoT settings, where intermittent participation and limited communication budgets make consistent performance and early stopping highly desirable.

\begin{figure}
    \centering
    \begin{minipage}{0.24\textwidth}
        \centering
        \includegraphics[width=\linewidth]{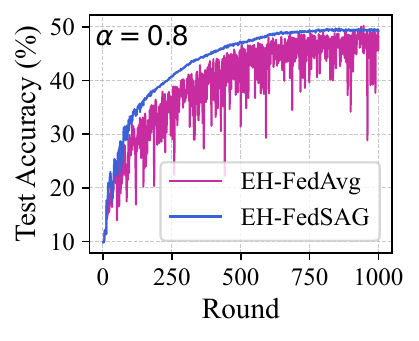}
    \end{minipage}
    \begin{minipage}{0.24\textwidth}
        \centering
        \includegraphics[width=\linewidth]{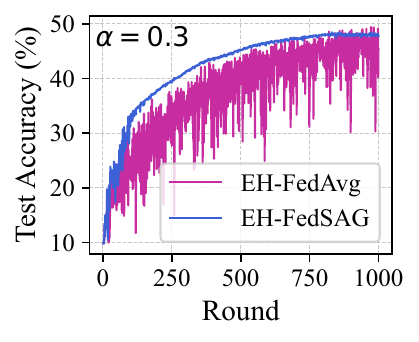}
    \end{minipage} 
    \caption{Test accuracy versus communication rounds associated with different Dirichlet coefficients ($\alpha$) for harvested probabilities $P_h=0.04$.}
\label{fig:accuracy_vs_round_alpha}
\end{figure}

To evaluate the performance of EH-FedSAG under non-i.i.d. training conditions, we partition the local datasets across devices according to a Dirichlet distribution with parameter $\alpha$, which controls the degree of statistical heterogeneity. As shown in Fig.~\ref{fig:accuracy_vs_round_alpha}, EH-FedAvg exhibits pronounced round-to-round fluctuations, and these fluctuations are more severe when $\alpha$ is smaller. Although increasing $\alpha$ from $0.3$ to $0.8$ slightly reduces the variance of EH-FedAvg, the method still suffers from frequent sharp accuracy drops. In contrast, EH-FedSAG maintains a much smoother accuracy trajectory with substantially smaller oscillations across the entire training process. The reason is that EH-FedAvg aggregates only the updates from the devices active in the current round. Under heterogeneous data distributions, this active subset may not adequately represent the global objective, which increases the round-to-round variability of the aggregated update. EH-FedSAG alleviates this issue by also exploiting the recently stored information from previously participating devices, making the server update less sensitive to the instantaneous set of active devices.
Fig.~\ref{fig:accuracy_vs_round_alpha} also shows that EH-FedSAG reaches its high-accuracy regime $23\%$ faster for $\alpha=0.8$ than for $\alpha=0.3$, and it achieves slightly higher test accuracy in the less heterogeneous setting. This behavior is expected because a larger $\alpha$ makes the local data distributions more similar across devices, so the local updates become better aligned and the global model improves more consistently across rounds.

\section{Conclusion}
In this paper, we studied FL in EH systems, where intermittent energy arrivals and limited uplink resources restrict device participation and communication efficiency. We proposed EH-FedSAG, a server-memory-based variance-reduced FL method inspired by SAGA, and evaluated it against EH-FedAvg. Simulation results demonstrated that EH-FedSAG consistently provides higher test accuracy and substantially lower training variance across the considered scenarios, with especially notable improvements for non-i.i.d. data distributions. Overall, the results suggest that incorporating variance reduction into EH-FL is an effective way to enhance training stability in the sense of producing more consistent and predictable model performance across communication rounds. Future work may extend this framework to unreliable wireless links, theoretical convergence analysis under EH constraints, and more realistic EH-FL systems with computationally heterogeneous devices.

\bibliographystyle{IEEEtran}
\bibliography{reference.bib}

\end{document}